# Limit cycles for speech

Adamantios I. Gafos[a,1] and Stephan R. Kuberski[a]

**Rhythmic fluctuations in acoustic energy and accompanying neuronal excitations in cortical oscillations are characteristic of human speech, yet whether a corresponding rhythmicity inheres in the articulatory movements that generate speech remains unclear. The received understanding of speech movements as discrete, goal-oriented actions struggles to make contact with the rhythmicity findings. In this work, we demonstrate that an unintuitive – but no less principled than the conventional – representation for discrete movements reveals a pervasive limit cycle organization and unlocks the recovery of previously inaccessible rhythmic structure underlying the motor activity of speech. These results help resolve a time-honored tension between the ubiquity of biological rhythmicity and discreteness in speech, the quintessential human higher function, by revealing a rhythmic organization at the most fundamental level of individual articulatory actions.**

speech movements | discreteness | limit cycles | rhythm

Extensive research across neuroscience and acoustics points to a low-frequency speech rhythm, evident in acoustic amplitude fluctuations and concomitant auditory-motor cortical oscillations (1). At the same time, discreteness, a hallmark of linguistic cognition (2), is taken as the defining property of speech. Discrete choice, phoneme after phoneme, breaks with rhythmicity at the level of individual speech effectors. This tension between rhythmicity and discreteness underlies a conceptual disjuncture in the received understanding of speech motor action. In recognition of the rhythmicity perspective, by far the most favored models of speech movements are in principle oscillatory (3–5). Yet, in recognition of also the discreteness perspective, they are parameterized to generate movements that do not oscillate: speech gestures are conventionally assigned point attractor dynamics with critical damping (6). Such systems lack an intrinsic oscillation property, a severe setback for the aim of characterizing the rhythmicity of speech using discrete building blocks. There exist no established results that link point attractor dynamics to the well-documented evidence for a low-frequency 2–8 Hz periodicity in acoustic amplitude studies and corresponding cortical oscillations (1) or to theoretical modeling of the motor cortex as an oscillator (7, 8). This has left a major gap in our understanding of how a continuous rhythmic activity can be forged from episodic, temporarily assembled dynamics driving the movements of a biomechanically diverse periphery.

## Results

The accepted wisdom by the speech production community asserts that actions in speech – such as forming and releasing constrictions at different places in the vocal tract – are governed by point attractor dynamics (6, 9, 10). The dynamics is an equation that specifies how articulatory states change over time and the point attractor is a state that trajectories approach asymptotically, i.e., an equilibrium. Central to this view is the assumption that the equilibrium $x_0$ is identified with the endpoint (offset) of the movement. For example, in forming the tongue-palate constriction for [k], the offset is when the tongue dorsum comes into full contact with the palate. Here we relax this commitment – equilibrium at movement offset – by taking the seemingly unintuitive

Author affiliations: [a]University of Potsdam, Germany

Both authors contributed equally to this work. Authors' names are listed alphabetically.

No author has any conflict of interests to declare.

[1]Corresponding author: gafos@uni-potsdam.de

conceptual leap of placing the equilibrium at different locations along a movement's trajectory. We chose five representative $x_0$ placements, expressed as percentages along the movement excursion: $x_0 = 100\%$ (equilibrium at offset–the classic view), $x_0 = 75\%$ (equilibrium at three-quarters of the excursion), $x_0 = 50\%$ (midpoint equilibrium), $x_0 = 25\%$ (equilibrium at one-quarter of the excursion), and $x_0 = 0\%$ (equilibrium at movement onset). Though it may seem intuitive to identify $x_0$ with movement offset, proponents of the dynamical approach to motor cognition have emphasized all along that the underlying organization governing movements is oscillatory (3). If so, another arguably natural choice is to set $x_0$ midway in the movement's excursion, much as in a pendulum where the equilibrium is midway between the two positional extremes. Whatever their apparent naturalness, neither this nor other choices like it have been empirically evaluated. Here, we resolve this issue by rigorously evaluating the consequences of different $x_0$ choices.

Nearly half a century ago, Kelso and colleagues (3–5) proposed that the informal concept of a multi-effector synergy, as typified in the movements that produce phonemes (11), corresponds to the formal concept of a limit cycle. Limit cycles are the mathematical nonlinear archetype of oscillatory behavior and have proven indispensable as organizing principles of temporal order in the natural world (12). Speech movements of varying phonemic identities and kinematic extents – from labial, coronal, or dorsal consonants to various vowels – were sourced from an extensive multi-archive articulatory dataset spanning American and British English (see Supplementary Information). Regressions in phase space were used to assess the fit between the movements and three archetypal limit cycle oscillators (van der Pol, Rayleigh, and hybrid; see Supplementary Information). In our assessment, we kept with the assumption that, for discrete movements, each opening (e.g., [k] to [a]) and each closing movement (e.g., [a] to [k]) corresponds to a half cycle of an underlying oscillatory dynamics (13), but changed the dynamics from linear point attractors to nonlinear limit cycles. Thus, a discrete movement, as in [k] to [a], would be a traversed arc in phase space on a limit cycle trajectory; the next movement would live on a different (re-parameterized) limit cycle, just as envisioned in the early eighties: "we might expect each cycle to be instituted de novo in speech to satisfy local phonetic [...] constraints" (4). Figure 1 (left) shows regression performances of the three oscillators in R-squared as a function of $x_0$ choice: all three oscillators peak at midpoint $x_0$ (50% between onset and offset). In short, revising the endpoint-as-equilibrium prevailing wisdom enables predicting the movements with substantially higher accuracy compared to the conventional view ($x_0$ at offset). This finding is not dependent on any one dynamical formulation within the class of the three prototypical limit cycle oscillators (van der Pol, Rayleigh, and hybrid). The choice of centering movements on an equilibrium, rather than anchoring them at their offset, was always available but so far disregarded due to conventional assumptions about the mapping between anatomical landmarks and the equilibrium of the underlying dynamics.

The same choice of centering movements on an equilibrium that substantially enhances the accuracy of predicting the kinematics compared to the conventional view has another immediate consequence: it enables the recovery of a long-sought rhythmicity underlying the kinematics of speech. Several lines of evidence, based on acoustic envelope periodicities, syllable rate, auditory and motor cortex oscillations, point to a low-frequency 2–8 Hz periodicity for speech. Within that broad range of two octaves, a narrower, dominant range between 4 and 5 Hz is observed in the power spectrum of the speech amplitude envelope. A thorough review is in ref. (1). The same narrow range (4–5 Hz) is thought to be privileged in the synchronization of speech-motor and auditory cortical areas during speech processing (8). It has been suggested that the converging evidence for low-frequency periodicities is a consequence of speech gestures being "executed at a relatively regular interval" (1). Long before rigorous evidence supported this idea, Wilhelm Wundt, proposed that rhythmic "motor pulses" throttle the flow of speech movements, suggesting that even non-repetitive speech has an underlying rhythmic structure (14). A cyclicity in vocal tract aperture is of course reminiscent of the familiar syllable structure in many languages (a consonant followed by a vowel and so on) and it is the acoustic consequences of these consonant-vowel alternations existing techniques have exploited to demonstrate a low-frequency mode in acoustic envelope periodicities (1). However, at the neural level, sensorimotor cortex activity during speaking is not known to encode gross aperture changes (that is, undifferentiated cycles of opening and closing) but rather phoneme-specific constrictions by anatomically separate articulators (15). An articulatory-motor foundation for cyclicity, if it exists, ought to reflect this specificity.

When on a limit cycle, a system oscillates with a characteristic frequency which is a function of the system's intrinsic parameters. Figure 1 (middle panel) shows limit cycle frequencies of the three nonlinear oscillators as a function of $x_0$ choice. These frequencies derive from the regressed oscillators to the same data. At midpoint $x_0$, all three oscillators yield mean frequencies between 4 and 5 Hz, thus matching the widely reported acoustic and cortical rhythms. As $x_0$ drifts away from midpoint, the match between the motor rhythms and the rhythms observed in acoustic and cortical dynamics progressively deteriorates. These frequency means are grand averages across all phonemes (labial, coronal, and dorsal consonants and vowels) and across all their contexts of occurrence in the data. Figure 1 (right panel) shows that recovered frequencies per place of articulation for consonants and vowel height for vowels fall squarely within a 2–8 Hz range. These frequencies are derived using the hybrid oscillator at midpoint $x_0$. Per place of articulation and vowel height frequency statistics show values around the 4–5 Hz range. In sum, a long-anticipated finding that had until now remained elusive, despite extensive converging evidence from different sources pointing to it, emerges: a typicality in the 4–5 Hz range is recovered directly from the motor activity at the level of specificity required for linguistically-relevant actions. This rhythmicity arises naturally when movements are generated from an intrinsically nonlinear cyclic organization, the same organization widely acknowledged across physics, biology, and neuroscience to underlie spatiotemporal order in the natural world.

Rhythms underlie many important behaviors in biology generally and in animal communication specifically. A better understanding of different species' vocal skill and rhythmic behavior is essential in testing hypotheses about the evolution of speech. In particular, rhythmic orofacial movements in ancestral primates are hypothesized precursors to speech (16). Notably, during lip-smacking, a face-to-face interactive gesture in macaques, lips and tongue also move at 4–5 Hz (17), consistent with the rhythm of movements identified here in speech.

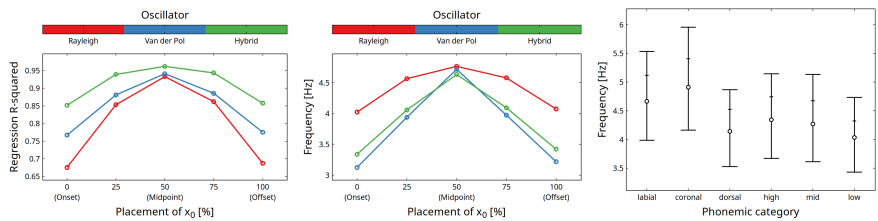

**Fig. 1. Left:** Regression R-squared of three non-linear limit cycle oscillators (van der Pol, Rayleigh, hybrid) for different choices of setting the equilibrium $x_0$ of the dynamics. For each oscillator, peak performance, near unity, is achieved at $x_0$ midpoint (50% between onset and offset), challenging the received view which has $x_0$ set at movement offset (100%). **Middle:** Frequencies of the three oscillators for different choices of $x_0$ across all phonemes. **Right:** Frequency statistics – median (circle), mean, first and third quartiles (dashes) – of the hybrid oscillator at midpoint $x_0$ per place of articulation (labial, coronal, dorsal consonants) and vowel height (high, mid, low vowels).

## Conclusion

Rhythmicity is ubiquitous in biological processes (12). Yet, especially in the context of more complex cognitive functions such as speech, the import of the fact that motor systems naturally generate rhythmic actions has remained unclear (18). In language, discrete infinity (2), the ability to create an infinite set of new forms from a finite set of discrete units, is considered a hallmark property, an exponent of the flexibility of human behavior and the generative capacity of the human mind. Discrete infinity has seemed at best unrelated to and at worst irreconcilable with speech's putative rhythmic motor foundations (18). Our findings point to an approach that transcends the rhythmicity-discreteness dichotomy, by demonstrating a long-sought rhythmicity at the most fundamental level of individual articulatory actions. In a companion work using cross-linguistic datasets, we demonstrate that the same results apply to speech movements from several typologically diverse languages.

Rhythmic activities forged from temporarily assembled dynamics that recruit a biomechanically diverse periphery of effectors abound in nature. Whether the same approach that uncovers a rhythmic organization in the articulatory-motor basis of speech also applies to other forms of action, from music and dance to the vocal behaviors of nonhuman animals, remains an open question with far-reaching implications for the study of communication and movement in general.

## Materials and Methods

Detailed information about materials and methods is given in the supplementary document.

**ACKNOWLEDGMENTS.** This work has been funded by the European Research Council under Grant Agreement ID 249440 and by the German Research Foundation (Deutsche Forschungsgemeinschaft) under Project ID 317633480, SFB 1287, Project C04. Appreciation is given to Dung Nguyen for his assistance in algorithm design, code implementation and other technical aspects. We are grateful to Mark Tiede for facilitating access to the HPRC-IEEE archive of American English and to Alan Wrench for communication on the MOCHA-TIMIT archive of British English.

# Supplementary information

Adamantios I. Gafos[a,1] and Stephan R. Kuberski[a]

## Oscillators

The van der Pol oscillator (1) is a classic example of a nonlinear oscillator that features limit cycle dynamics. Its governing equation reads $\ddot{x} = -x + \mu(1 - x^2)\dot{x}$ with the damping ratio $\mu$ controlling the properties of the limit cycle; hence, we have a one-parameter family of functions $f(x, \dot{x})$, parameterized by $\mu$. The Rayleigh oscillator (2) is another classic example of a limit cycle oscillator. Its equation reads $\ddot{x} = -x - \nu(1 - \dot{x}^2)\dot{x}$, with the damping ratio $\nu$ similarly controlling the properties of the limit cycle. The two oscillators exemplify two canonical nonlinear damping mechanisms: $\mu(1 - x^2)\dot{x}$, identified by van der Pol in the early 20th century, and $\nu(1 - \dot{x}^2)\dot{x}$, identified by Rayleigh in the late 19th century, both yielding limit cycle behavior in a range of physical and biological systems. Human movement patterns may recruit these mechanisms in combination. Hence, in broadening the class of limit cycle behaviors rendered by the two classic oscillators, the hybrid oscillator (3) combines their equations to $\ddot{x} = -x + \mu(1 - x^2)\dot{x} + \nu(1 - \dot{x}^2)\dot{x}$.

A time-honored model for speech gestures is the critically damped linear oscillator (CDO) $\ddot{x} = -x - 2\dot{x}$ (4). We evaluated CDO using the same methods (see below) as for the three limit cycle oscillators. Its regression performance (R-squared values) lies in the range of 0.001 (at onset $x_0$) and 0.4 (at offset $x_0$). In other words, CDO's best performance is worse than the worst performance of any limit cycle oscillator. This result is unsurprising. Although CDO's linear dynamics have provided a proof-of-concept model for an oscillatory basis of speech, there is by now ample evidence of difficulties with empirical coverage and hints for the existence of nonlinearities in speech movement trajectories. Furthermore, frequencies estimated from the CDO reach a maximum of 3.5 Hz and thus align poorly with the low-frequency 2–8 Hz range and in particular fall outside the dominant 4–5 Hz band. Accordingly, we omit CDO regression and frequency metrics from the main text.

## Materials

Speech movements of varying phonemic identities and kinematic extents (labial, coronal or dorsal consonants to various vowels) were selected for analysis. They derived from the electromagnetic articulography (EMA) channels of the Multichannel Articulatory (MOCHA–TIMIT) database (5), the EMA Haskins Production Rate Comparison (HPRC–IEEE) database (6), and the X–Ray Microbeam (XRMB) database (7). MOCHA–TIMIT includes speech movement recordings from two British English speakers (one female, one male) at a sampling rate of 500 Hz. The speakers read 460 sentences sourced from the TIMIT corpus. HPRC–IEEE consists in speech movement recordings from eight speakers (four females, four males) at a sampling rate of 100 Hz. The

Author affiliations: [a]University of Potsdam, Germany

Both authors contributed equally to this work. Authors' names are listed alphabetically.

No author has any conflict of interests to declare.

[1]Corresponding author: gafos@uni-potsdam.de

speakers read 720 sentences from the IEEE corpus. Finally, XRMB includes speech movement recordings from 57 American English speakers (32 female, 25 male) at a sampling rate of 146 Hz. We used a subset of 48 speakers. The speakers produced speech in a variety of tasks including the production of single sentences, entire passages and word lists.

The total number of phonemes from which kinematics was extracted across all datasets amounts to (in thousands): 76 labial (/p, b, m, f, v, w/), 249 coronal (/t, d, n, θ, ð, s, z, ʃ, ʒ, tʃ, dʒ, l, r/) and 38 dorsal consonants (/k, g, ŋ, j/), as well as 56 high (/i, ɪ, u, ʊ/), 101 mid (/e, ɛ, ɝ, ə, ʌ, o, ɔ, oʊ/), and 24 low vowels (/a, æ/). Each phoneme contributes a closing and an opening movement of the primary effector associated with that phoneme (lips for labial, tongue tip for coronal, tongue dorsum for dorsal consonants, and tongue dorsum for vowels).

## Methods

The two-dimensional (MOCHA-TIMIT and XRMB) or three-dimensional (HPRC-IEEE) effector trajectories were segmented using classic velocity-based methods, where a segment is the period of effector motion starting at one velocity minimum (movement onset), crossing a maximum (peak velocity), and ending at another, next velocity minimum (movement offset). The two velocity minima of each segment served to define the local principal direction of motion in body space, the so-called reaching axis (8), oriented along the line from initial position of the effector to the constriction target. Movement trajectories were then projected onto their principal direction. This projection constitutes a transformation from a representation in body-spatial (anatomical) coordinates to a representation in functional coordinates, expressing the task of forming or releasing a constriction regardless of where in body space the action unfolds.

Polynomial regressions were finally used to assess the fit between the so-obtained segmented representations of movements and the hypothesized oscillators. From the fitted models, we derived the limit cycle parameter of frequency and computed mean values using pooled statistics along the relevant dimensions in our datasets, e.g., speaker identity and phonemic identity. Model fits and frequency estimates reported in the main text are pooled across the British English (MOCHA-TIMIT) and American English (HPRC-IEEE and XRMB) datasets. When treating British and American English varieties individually, the per-variety regression performance of the three oscillators are as follows (in R-squared at midpoint $x_0$): 0.94 (van der Pol), 0.94 (Rayleigh), 0.96 (hybrid) for British English; 0.94 (van der Pol), 0.93 (Rayleigh), 0.96 (hybrid) for American English. The recovered frequencies, also derived at midpoint $x_0$, are 4.6 Hz (van der Pol), 4.6 Hz (Rayleigh), 4.5 Hz (hybrid) for British English; 4.7 Hz (van der Pol), 4.8 Hz (Rayleigh), 4.6 Hz (hybrid) for American English.